\documentclass[12pt,useAMS,useAASTEX,usenatbib]{aastex}
\usepackage{emulateapj5}

\makeatletter
\newenvironment{inlinetable}{%
\def\@captype{table}%
\noindent\begin{minipage}{0.98\linewidth}\begin{center}\footnotesize}
{\end{center}\end{minipage}}%\smallskip}

\newenvironment{apjemufigure}{%
\def\@captype{figure}%
\noindent\begin{minipage}{0.999\linewidth}\begin{center}}
{\end{center}\end{minipage}}%\smallskip}
\makeatother

\def\healpix{H{\sc ealpix }}

\def\etal{et al.}

\def\alm{a_{\ell m}}
\def\Ylm{Y_{\ell m}}
\def\Cl{C_{\ell}}

\def\summ{\sum_{m=-\ell}^{\ell}}
\def\suml{\sum_{\ell=0}^{\infty}}

\newcommand{\tac}{{Theoretical Astrophysics Center, Juliane Maries Vej
30, DK-2100,  Copenhagen, Denmark}}
\newcommand{\nbi}{{Niels Bohr Institute, Blegdamsvej 17,
DK-2100 Copenhagen, Denmark}}
\newcommand{\sao}{{Special Astrophysical Observatory, Nizhnij Arkhyz,
Karachaj-Cherkesia, 369167, Russia}}

\newcommand{\ru}{{Rostov State University, Space Research Department,
Zorge,5, 344091, Russia}}

\slugcomment{Submitted to {\it The Astrophysical Journal Letters}}

\begin{document}

\title{Phase cross-correlation of the {\it WMAP} ILC map and foregrounds}

\author{
Pavel D. Naselsky\altaffilmark{1-3},
Andrey G. Doroshkevich\altaffilmark{1},
Oleg V. Verkhodanov\altaffilmark{1,4},
%Lung-Yhi Chiang\altaffilmark{1},
%Igor Novikov\altaffilmark{1,2}
}

\altaffiltext{1}{\tac}
\altaffiltext{2}{\nbi}
\altaffiltext{3}{\ru}
\altaffiltext{4}{\sao}

\email{ naselsky@tac.dk, dorr@tac.dk, oleg@tac.dk}

\begin{abstract}
 
We present a circular cross-correlation tests 
for the phases  
of the Internal Linear Combination Map (ILC)  and {\it WMAP}'s foregrounds 
 for all K--W frequency bands
 at the range of multipoles $\ell\le 50$.
We have found significant deviations from the expected Poissonian
statistics for the ILC and the foregrounds phases.
Our analysis shows that the low multipole range of the ILC power spectrum
contains some of the foregrounds residues.

\end{abstract}

\keywords{cosmology: cosmic microwave background --- cosmology:
observations --- methods: data analysis}

\section{Introduction}

The recently-published Wilkinson Microwave Anisotropy Probe
({\it WMAP}) data sets (see Bennett et al. 2003 a-c, Hinshaw 
et al. 2003 a-b)) strongly promote the development of the
high sensitive statistics for testing of the properties of the
derived foregrounds and foreground cleaned maps (Komatsu et al. (2003),
Tegmark, de Oliveira-Costa and Hamilton (2003), Chiang et al. (2003),
Dineen and Coles (2003), Park (2003), Gaztanaga et al. (2003),
Colley and Gott (2003), Naselsky et al. (2003)).
The {\it WMAP} team produces
the Internal Linear Combination (ILC) map and  maps for the foregrounds
(synchrotron, free-free, dust emission) for each K--W bands which are the
basis for our analysis. Homogeneous and isotropic CMB 
Gaussian random fields, as a result
of the simplest inflation paradigm, is a crucial test for the early
stages of the cosmological expansion.
Because of the pronounced
non-Gaussianity of the foregrounds it would be
natural to expect that possible non-Gaussianity of the ILC and
other cleaned maps derived from the {\it WMAP} data
would reflect directly contamination of the foregrounds
at different levels. 

In this Letter we present the result of a statistical test of
the coupling the ILC map and the foreground taking from the
{\it WMAP} data\footnote{see \tt http://lambda.gsfc.nasa.gov}.
The test is based on the circular
cross-correlations analysis of the
ILC and foregrounds phases. 
Our method  complementing the mentioned above ones,
exploits a natural assumption that
phases of the ``true'' CMB signal  should not correlate
with the phases of the
foregrounds\footnote{Naselsky et al. 2003 have shown that
for any linear separations of the CMB signal and foregrounds
the phases of derived CMB signal should be correlate with
the foreground phases at different level depending on
separation technic.}.
This allows us
to detect significant ILC--``W band foregrounds'' cross-correlations.
These peculiarities determine the accuracy of the power spectrum
estimation for the ILC map and the statistical properties of the ILC signal.
We would like to point out that the detected non-Gaussianity of the ILC map
(see for example, Eriksen et al. 2003, Vielva et al. 2003,
Coles et al. 2003) is most likely
 related 
with foreground contaminations (see also Chiang et al. 2003)
produced most likely by the Galactic dust emission.

\section{ ILC and foregrounds phase correlations}

The basic idea of the phase analysis of the CMB maps was proposed
and discussed
in Chiang and Coles (2000), Naselsky, Novikov and Silk (2002),
Chiang et al. (2002), Chiang et al. (2003), Naselsky et al. (2003),
Matsubara (2003)

The CMB signal from
can be expressed as
a sum over spherical
harmonics:
\begin{eqnarray}
\Delta T(\theta,\varphi)=\suml \summ \alm \Ylm (\theta,\varphi), \nonumber \\
\alm=|\alm|\exp(i\Psi_{\ell m}),
\label{eq1}
\end{eqnarray}
where $\alm $ are the coefficients of expansion, $|\alm|$ is
the modulus and $\Psi_{\ell m}$ is the phase of each $\alm$
harmonic.
Homogeneous and isotropic CMB Gaussian random fields (GRFs), as a  result
of the simplest inflation paradigm, possess Fourier modes whose real
and imaginary parts are independently distributed. The statistical
properties of GRF are completely specified by its angular power spectrum
$\Cl$,
\begin{eqnarray}
C(\ell,\ell',m,m')=\langle\alm a^{*}_{\ell'm'}\rangle= & \hspace*{1cm}\nonumber \\
\langle|\alm||a_{\ell'm'}
|\rangle\langle  \exp\left(i(\Psi_{\ell m}-\Psi_{\ell'm'})\right)\rangle=
     C(\ell)\delta_{\ell^{ } \ell^{'}} \delta_{m^{} m^{'}}  &
\label{eq4}
\end{eqnarray}
where $\langle..\rangle$ means averaging over the modulus and
phase distribution
functions.
In Fig.1 we plot the cross-correlation of phases between the ILC map
and the
common foregrounds
defined as a sum over synchrotron, free-free and dust emission maps 
for each frequency
band\footnote{Below we will use the term foreground for such
a combinations of each component.}.

\begin{apjemufigure}
\hbox{\hspace*{-0.5cm}
\centerline{
\includegraphics[width=0.52\linewidth]{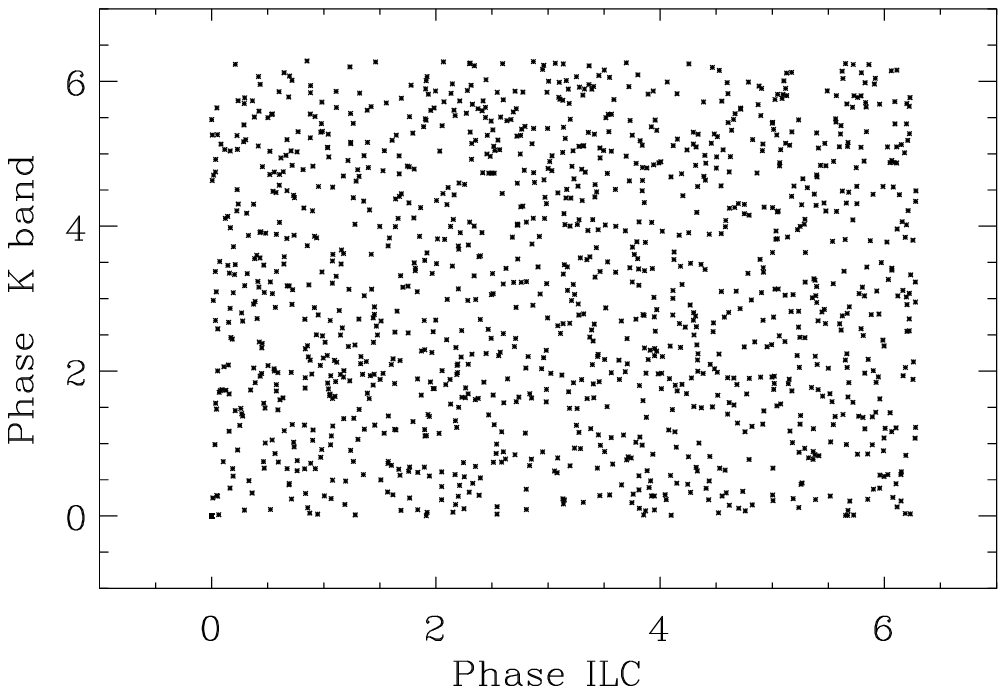}
\includegraphics[width=0.52\linewidth]{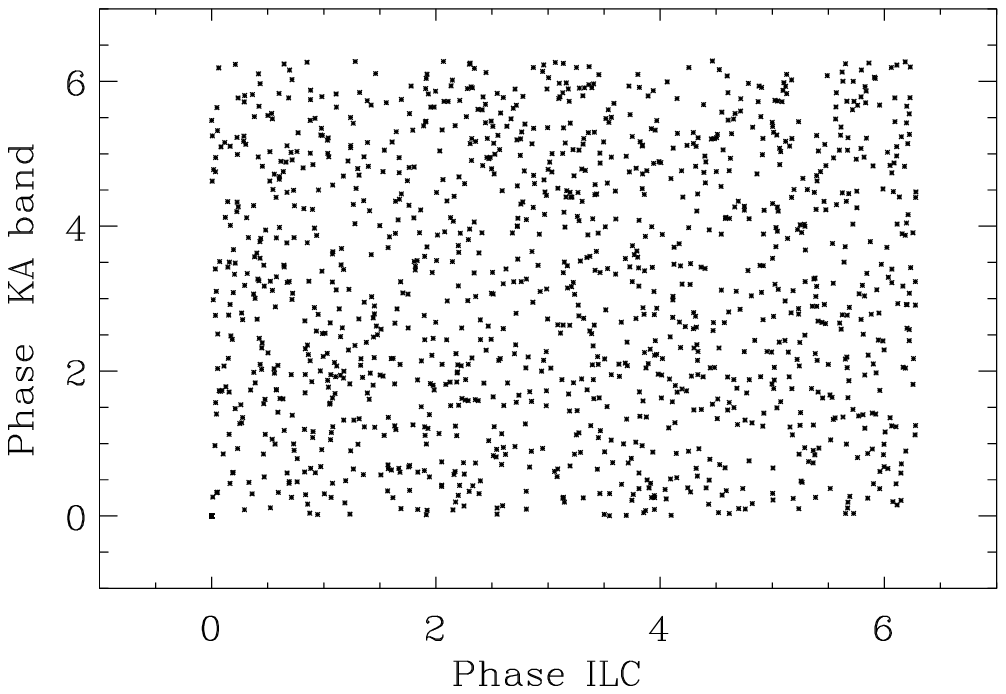}}}
\hbox{\hspace*{-0.5cm}
\centerline{
\includegraphics[width=0.52\linewidth]{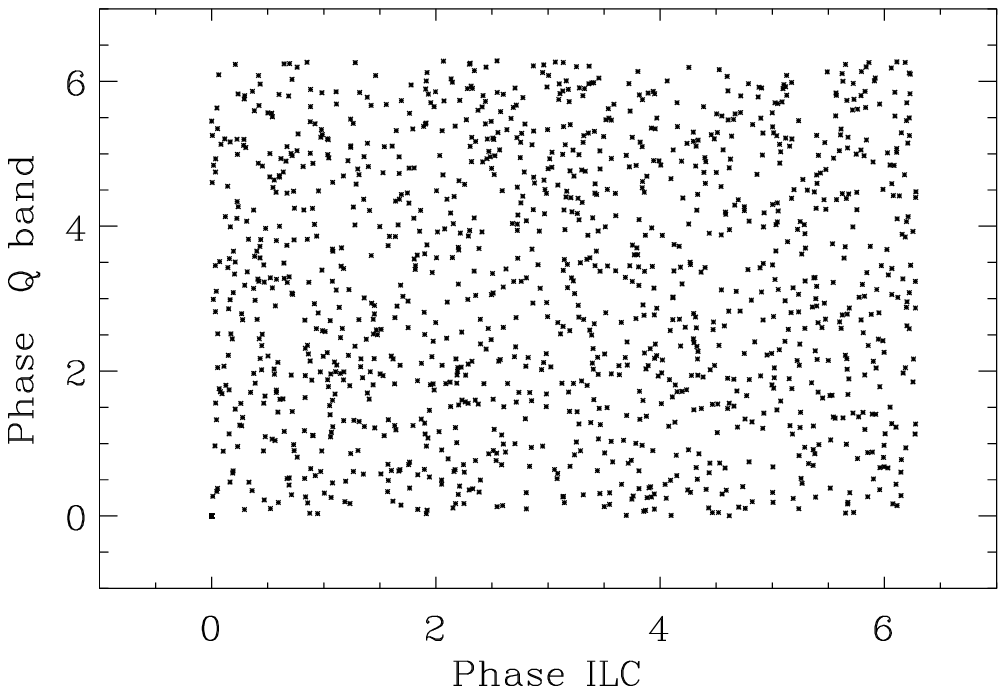}
\includegraphics[width=0.52\linewidth]{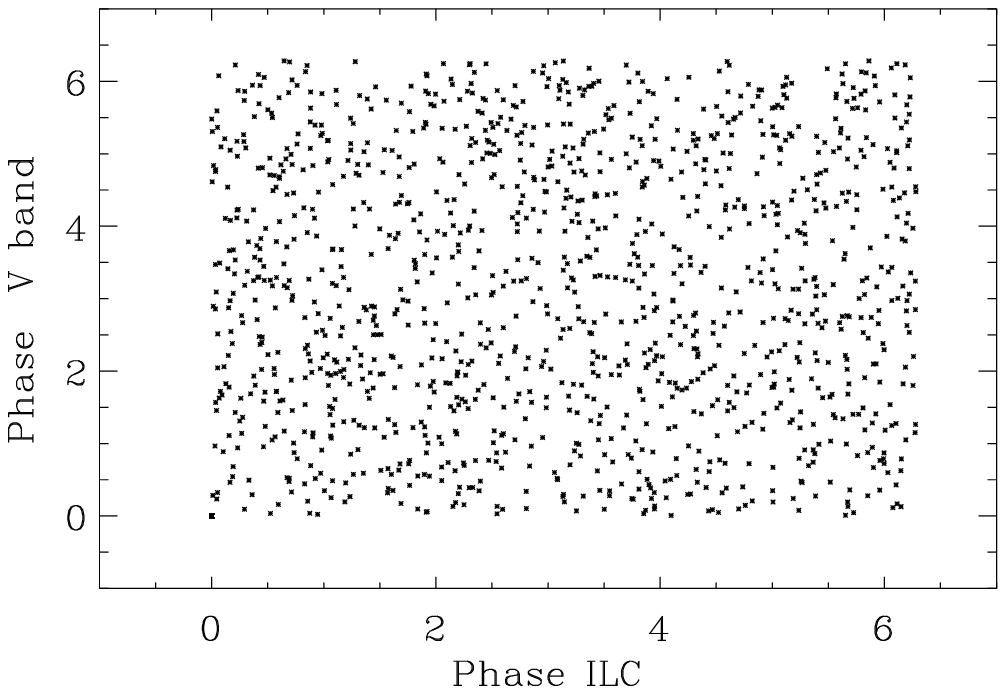}}}
\hbox{\hspace*{-0.5cm}
\centerline{
\includegraphics[width=0.52\linewidth]{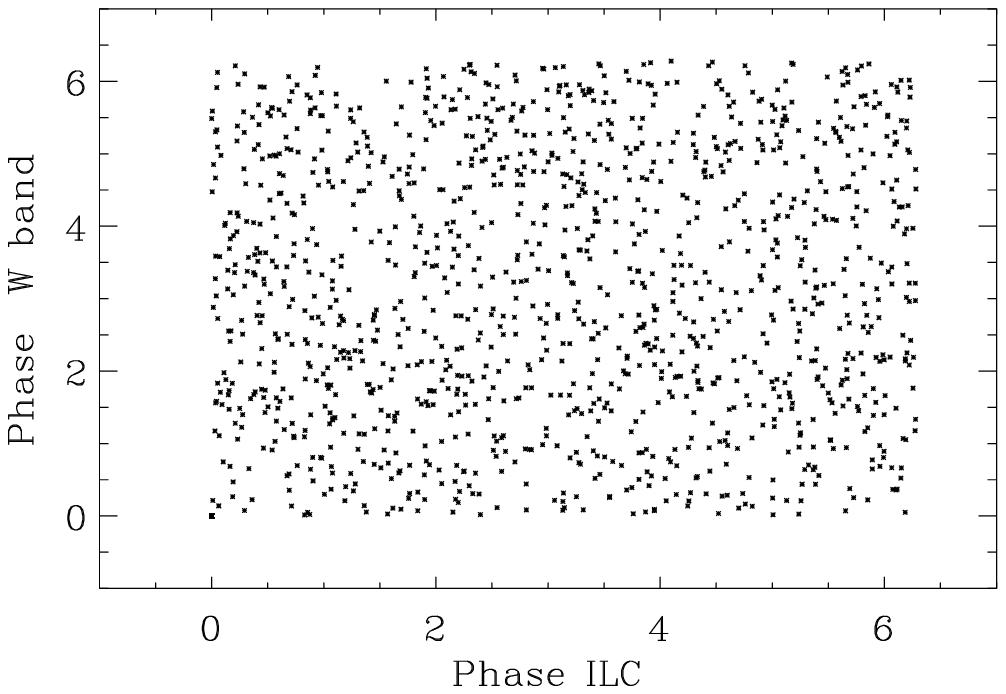}}}
\caption{The cross-correlation of phases for ILC and the K--W
bands foregrounds. All the foreground maps and
derived phases were taking from the {\it WMAP} web-site. } 
\label{fig1} 
\end{apjemufigure}

As one can see from Fig.1 the distribution of the ILC map phases 
and corresponding foregrounds
seems to be non-correlated one. However, in order to test such 
a distribution, 
for these phase diagram we will apply circular statistics (Fisher 1993)
to investigate the possible cross-correlation quantitatively.

\section{Correlations of the ILC and foregrounds phases}

Naselsky et al. (2003) have shown that for linear methods of the
foreground separation from the CMB signal we can
expect to find
some
cross--correlations
between the phases of the cleaned signal, $\phi_s$, and
the foregrounds, $\psi_f$, which can be used to characterize the degree
of separation achieved. To do this, here we will use
the circular correlation coefficients which reflect directly
the $2\pi$-phase periodicity.

For the K--W  frequency band the maps were taking from the
{\it WMAP} web
site\footnote{\tt http://lambda.gsfc.nasa.gov/product/map/m\_products.cfm}
and all the phases
are obtained by the spherical harmonic decomposition
using the HEALPix (G\'orski et al. 1999) and
the GLESP (Doroshkevich et al. 2003)
codes. We consider the ranges of multipoles, 
$2\leq\ell\leq\ell_{max}=50$ for which the {\it WMAP} own foreground
are presented in
the {\it WMAP} web site. In addition to the {\it WMAP} foregrounds
for each band
we produce the five maps as a difference between signal ($S$) in the
band and the ILC signal: $F=S-ILC$.
We will call the $F$-map as the derived foreground.
Let $\psi$ and $\phi$ be the foreground and ILC phases, respectively,
for given value
of $\ell$ and
all the corresponding values of $m$.
Following Fisher~(1993) we define the statistics
\begin{eqnarray}
\hspace*{-0.5cm}
x_m=\cos(\psi_m),   \hspace{0.2 cm} y_m=\sin(\psi_m),         & \hspace*{1cm} \nonumber\\
%                   \hspace{0.2 cm}
\nu_m=\cos(\phi_m), \hspace{0.2 cm}\mu_m=\sin(\phi_m),        & \hspace*{1cm} \nonumber\\
% \hspace*{-0.5cm}
M_p=(1/\ell_{max})\sum_m^{\ell_{max}}
     \exp\left(ip(\phi_m-\langle\phi\rangle)\right),          & \hspace*{1cm} \nonumber\\
%                 \hspace{0.5 cm}
\langle\phi\rangle=\tan^{-1}(\frac{\sum_m\mu_m}{\sum_m\nu_m}),& \hspace*{1cm} \nonumber\\
%\hspace*{-0.5cm}
M1_p=(1/\ell_{max})\sum_m^{\ell_{max}}
     \exp\left(ip(\psi_m-\langle\psi\rangle)\right),          & \hspace*{1cm} \nonumber\\
%                  \hspace{0.2 cm}
\langle \psi \rangle=\tan^{-1}(\frac{\sum_m y_m}{\sum_m x_m}),& \hspace*{1cm} \nonumber\\
R_{sf}(\ell)=\ell^{-1}\sum_{m=1}^\ell \cos(\phi_m-\psi_m),    & \hspace*{1cm} \nonumber\\
%                  \hspace{0.2 cm}
r_{sf}=(\ell_{max}-\ell+1)^{-1}\sum_\ell^{\ell_{max}} R_{sf}(\ell) & \hspace*{1cm}
\label{circ}
\end{eqnarray}
 where $M_p$ and $M1_p$ are the $p$-th trigonometric moments of the
phase samples,
$\langle \phi \rangle$ and
$\langle \psi \rangle$ are the corresponding mean directions of the samples,
$R_{sf}(\ell)$ is the circular cross-correlation coefficient in each mode
$\ell$ and $r_{sf}$ is the mean circular cross-correlation coefficient
for all phases.
For 
$m=0$ and for all $\ell$ phases $\phi(\ell,0)=\psi(\ell,0)=0$
and here we neglect them.

For the K--W bands the
circular coefficients, $R_{sf}(\ell)$ are plotted in Fig. 2
for the ILC and its own foreground,
and for the ILC and the derived foreground.

\begin{apjemufigure}
\vspace{0.5cm}
\centerline{\includegraphics[width=0.8\linewidth]{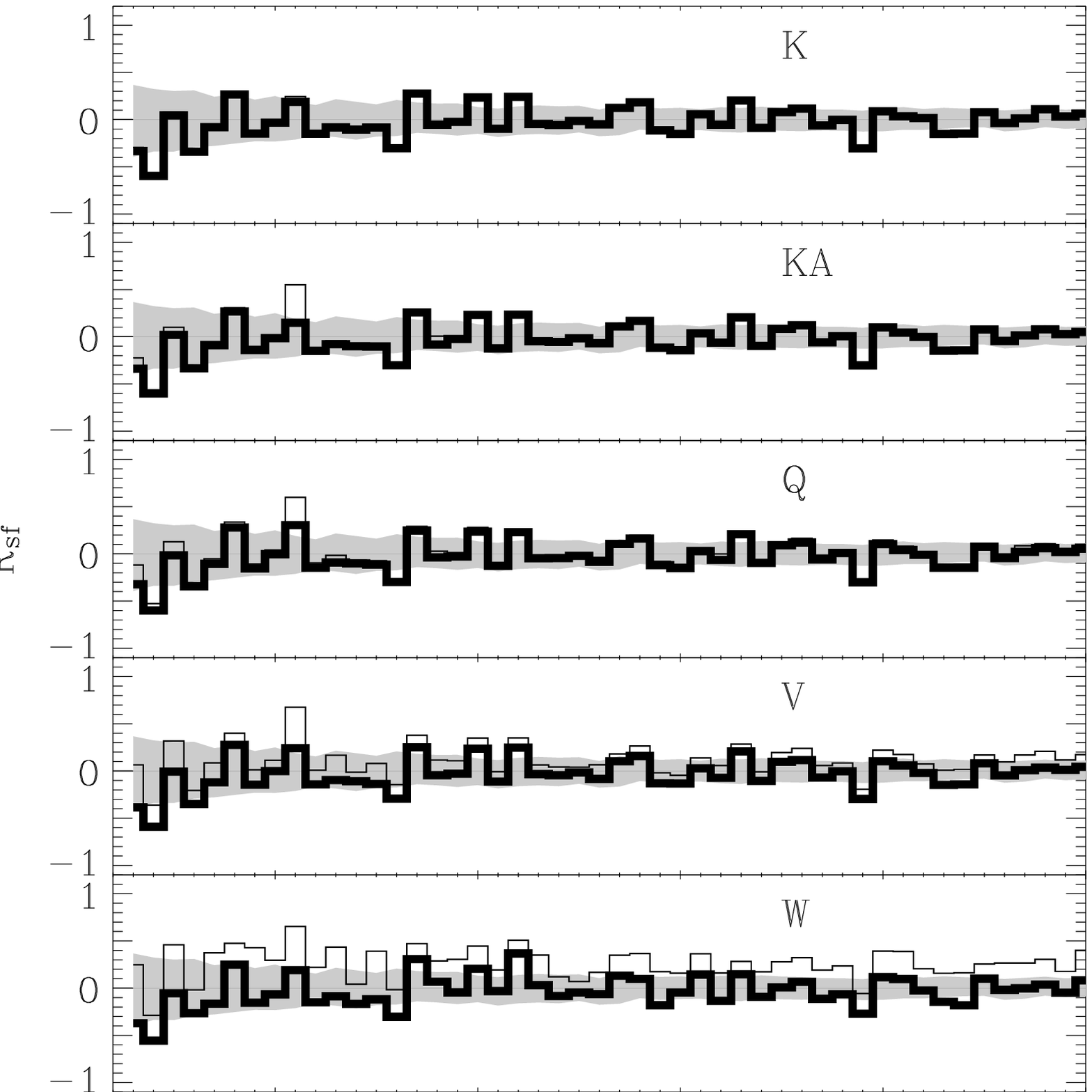}}
\vspace{0.6cm}
\caption{The circular correlation coefficient between the
cleaned signal and the foreground phases in V--W channels
vs. the harmonic index, $\ell$. Thick solid line corresponds to
the {\it WMAP} own foreground.
Thin solid line corresponds to derived foreground.
Shadow area represent $1\sigma$ error
bars level taking from 200 random realizations.
} 
\label{fc} 
\end{apjemufigure}

As is seen from the Fig.\,2,
for the first
three channels these coefficients are quite moderate and do not 
exceed the random scatter ($1\sigma$) obtained from 200 random 
realizations. 
Note that, 
for all the bands, the shape of the functions $R_{sf}(\ell)$
are quite similar to each other
which reflects the strong correlation
of phases in all the channels (Naselsky et al. 2003).
As one can see from Fig.2, the cross-correlation of the ILC phases and
the derived foreground seems to be more significant.
The same tendency follows from the
estimation of the mean
coefficients, $ r_{sf}$,
for $2\leq\ell\leq 50$, listed in Table\,1 for the two samples of
the foregrounds.
 As is expected, for the first three 
channels these coefficients are small but they become  
significant for channels V and W.

\vspace*{0.5cm}
\begin{inlinetable}
\caption{{\rm\small Circular cross-correlation
coefficients between phases of ILC and the {\it WMAP} own foregrounds 
(ILC$^{(o)}$) and the  ILC and derived foregrounds (ILC$^{(d)}$)}.}
\begin{tabular}{l ccc ccc} %7c
\hline
&K&KA&Q&V&W\cr
\hline
ILC$^{(o)}$  & -0.026 &-0.031 &-0.030 &-0.033 &-0.033 \cr
ILC$^{(d)}$  & -0.017 & 0.002 & 0.022 & 0.112 & 0.262 \cr
\hline
\end{tabular}
\label{Tab11}
\end{inlinetable}
\vspace*{0.5cm}

To test the sensitivity of circular statistics, we
used the so called ``WMAP simulator'', in which we simulated the W band
with
all own {\it WMAP} foregrounds
and the CMB signal with random phases
instead of the ILC signal\footnote{we
used the best fit {\it WMAP} $\Lambda$CDM cosmological
model from CMBfast by Seljak and
Zaldarriaga (1996).}. After that we repeat the circular  phase analysis
in the same way as it was
done for the ILC map. Our results are shown in Fig.\,3.
  
\vspace*{0.5cm}
\begin{apjemufigure}
\hbox{\hspace*{-0.8cm}
\centerline{\includegraphics[width=0.8\linewidth]{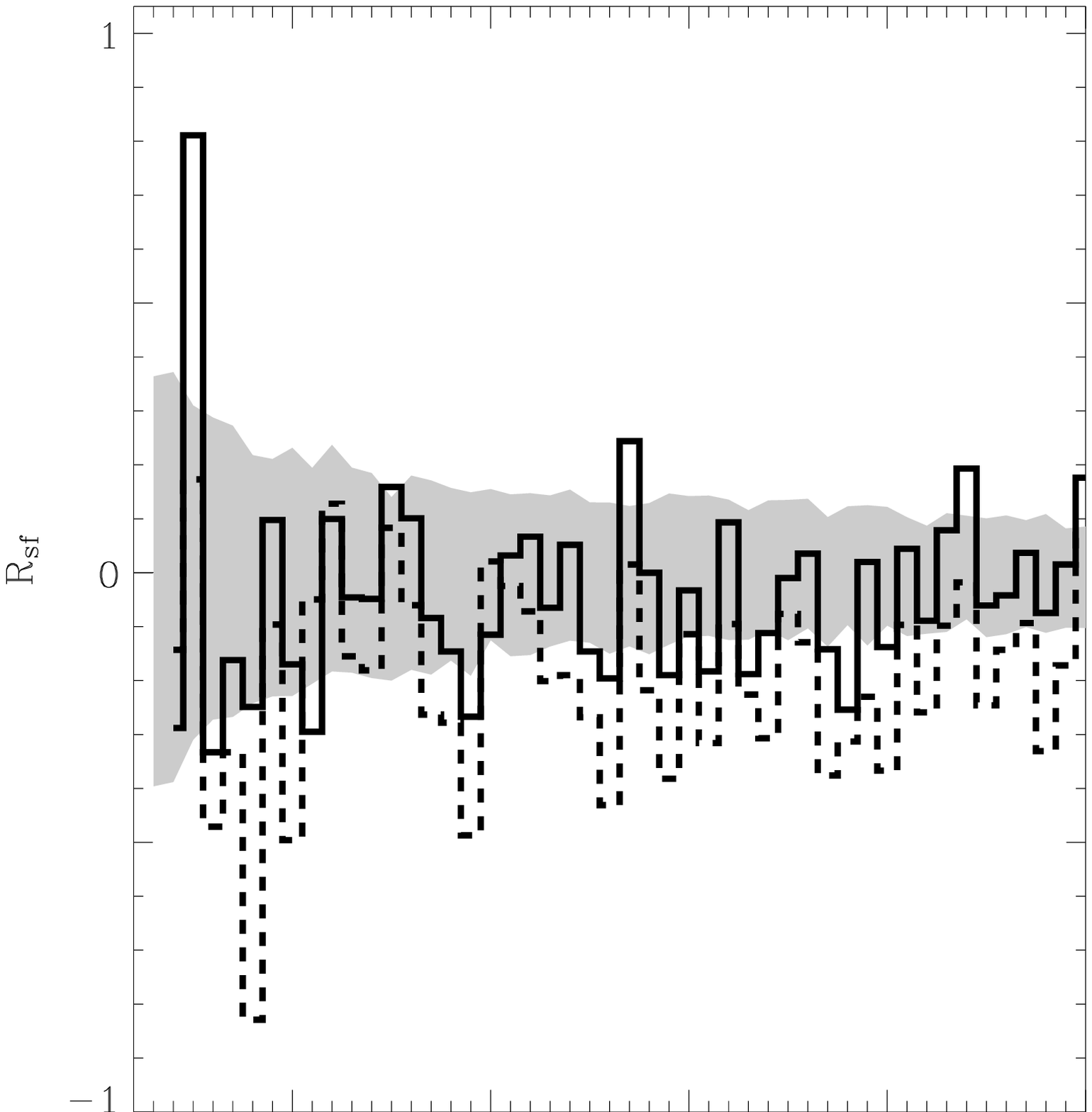}}}
\vspace*{0.6cm}
\caption{$R_{sf}$ for the ``WMAP simulator''.
Thick solid line corresponds to the simulated random CMB signal,
cross-correlated with {\it WMAP} own foreground.
For the CMB and derived foregrounds we have the same
curve. The dashed line is the cross-correlation between
the foreground, defined as the W band signal
minus the simulated CMB signal (not the ILC~!), and
the simulated CMB signal.
 } 
\label{fig3} 
\end{apjemufigure}

For the simulator, corresponding mean cross-correlation coefficients
are $r_{sf}=-0.036$ for
the CMB--Foreground correlations and $r_{sf}=-0.194$
for ``wrong''
foreground (actual signal minus simulated CMB), corresponds to
the dashed line in Fig.\,3.
Thus, if the CMB map and foreground are extracted correctly, corresponding
mean  cross-correlation coefficient should be less then statistical
error ($\sim 5\%$), while
the wrong foreground and CMB should have significant positive
or negative bias.
For the ILC map
it means that a positive bias for $r_{sf}$ for the
V and W bands (see Table\,1)
reflects directly the
cross-correlation with the foreground phases.

\section{Conclusion}

The circular cross-correlation analysis for the ILC map and K--W
foregrounds and
shows that the ILC map has significant correlation with
the derived foregrounds obtained by
subtraction the ILC map from
the V and W bands signals.
Some of this correlations
could be linked with non-Gaussianity of the ILC map detected
by Park (2003), Eriksen et al. (2003), Coles et al. (2003),
Vielva et al. (2003).
For example, Vielva et al. (2003) has reported about characteristic scale
of non-Gaussianity
in order to $10^\circ$, which corresponds to $\ell\sim 10$.
Our analysis shows that the cross-correlation
coefficients  $r_{sf}$ at $\ell=11$ have the highest maxima for both V
and W bands.
At high multipole range $\ell\sim 40$ corresponding peculiarities
are above $95\%$ confidential level.
If we take into account that for $\ell\sim 40$ all $m\sim 35-40$
lies along Galactic plane in the map,
 we can conclude that they are related with the V and W bands signal
at the same longitude.
Some of the peculiarities need an additional investigation
which would be published soon.

\section*{Acknowledgments}
This paper was supported by Danmarks Grundforskningsfond
through its support for the establishment of the Theoretical
Astrophysics Center. We thank Max Tegmark et al. for providing
their processed maps and making them public with openness.
We thank Igor Novikov and Lung-Yih Chiang
for useful discussions.
We also acknowledge the use of
\healpix\footnote{\tt http://www.eso.org/science/healpix/}
package \citep{healpix} to produce $\alm$ from {\it WMAP} maps
the use of GLESP package of the TAC CMB  collaboration.


\begin{thebibliography}{}
\expandafter\ifx\csname natexlab\endcsname\relax\def\natexlab#1{#1}\fi
\newcommand{\combib}[3]{\bibitem[{#1}({#2})]{#3}} %apj
%
% for authors
%
\newcommand{\autetal}[2]{{#1,\ #2., et al.,}}
\newcommand{\aut}[2]{{#1,\ #2.,}}
\newcommand{\saut}[2]{{#1,\ #2.,}}
\newcommand{\laut}[2]{{#1,\ #2.,}}
%
% for papers
%
% reference for papers: 1title, 2journal, 3vol, 4page, 5year, 6astro-ph
\newcommand{\refs}[6]{#5, #2, #3, #4} %apj
\newcommand{\unrefs}[6]{#5, #2 #3 #4 (#6)}  %apj
%
% for books and proceedings
%
% reference for books: 1title, 2press, 3editor, 4edition, 5year,
% 6astro-ph 
\newcommand{\book}[6]{#5, #1, #2} %mnras
%
% reference for books: 1title, 2press, 3editors, 4proc, 5year,
% 6astro-ph 
\newcommand{\proceeding}[6]{#5, in #3, #4, #2} %mnras  

\def\apj{ApJ}
\def\apjl{ApJL}
\def\mn{MNRAS}  
\def\nature{Nature} 
\def\aa{A\&A}   
\def\aaa{A\&A}
\def\prl{Phys.\ Rev.\ Lett.}
\def\prd{Phys.\ Rev.\ D}
\def\pr{Phys.\ Rep.}
\def\cambridgepress{Cambridge University Press, Cambridge, UK} 
\def\princetonpress{Princeton University Press}
\def\worldpress{World Scientific, Singapore}
\def\oxfordpress{Oxford University Press}

\combib{Bennett~et al.}{2003a}{wmap}
\autetal{Bennett}{C. L}
% Bennett, C. L.; Bay, M.; Halpern, M.; Hinshaw, G.; Jackson, C.;
% Jarosik, N.; Kogut, A.; Limon, M.; Meyer, S. S.; Page, L.; Spergel, D. N.;
% Tucker, G. S.; Wilkinson, D. T.; Wollack, E.; Wright, E. L.
\refs{}
{\apj}
{583}
{1}
{2003}
{}

\combib{Bennett~et al.}{2003b}{wmapb}
\autetal{Bennett}{C. L}
%  Bennett, C. L.; Halpern, M.; Hinshaw, G.; Jarosik, N.; Kogut, A.;
%  Limon, M.; Meyer, S. S.; Page, L.; Spergel, D. N.; Tucker, G. S.;
%  Wollack, E.; Wright, E. L.; Barnes, C.; Greason, M. R.; Hill, R. S.;
%  Komatsu, E.; Nolta, M. R.; Odegard, N.; Peiris, H. V.; Verde, L.;
%  Weiland, J. L.
\refs{}
{\apjs}
{148}
{1}
{2003}
{astro-ph/020320}

\combib{Bennett~et al.}{2003c}{wmapfg}
\autetal{Bennett}{C. L}
% Bennett, C. L.; Hill, R. S.; Hinshaw, G.; Nolta, M. R.; Odegard, N.;
% Page, L.; Spergel, D. N.; Weiland, J. L.; Wright, E. L.; Halpern, M.;
% Jarosik, N.; Kogut, A.; Limon, M.; Meyer, S. S.; Tucker, G. S.; Wollack, E.
\refs{}
{\apjs}
{148}
{97}
{2003}
{astro-ph/0203208}

\combib{Chiang \& Coles}{2000}{}
\aut{Chiang}{L.-Y} \laut{Coles}{P}
\refs{}
{\mn}
{311}
{809}
{2000}
{}

\combib{Chiang~\etal}{2002}{}
\aut{Chiang}{L.-Y} \aut{Coles}{P} \laut{Naselsky}{P. D}
\refs{}
{\mn}
{337}
{488}
{2002}
{}

\combib{Chiang~\etal}{2003}{}
\aut{Chiang}{L.-Y} \aut{Naselsky}{P. D}
\aut{Verkhodanov}{O. V} \laut{Way}{M. J}
\refs{}
{\apjl}
{590}
{65}
{2003}
{}

\combib{Coles\etal}{2003}{}
\aut{Coles}{P} \aut{Dineen}{P} \aut{Earl}{J} \laut{Wright}{D}
\unrefs{}
{\mn}
{submitted}
{}
{2003}
{astro-ph/0310252}

\combib{Colley \& Gott }{2003}{}
\aut{Colley}{W. N} \laut{Gott}{J. R}
\refs{}
{\mn}
{344}
{686}
{2003}
{astro-ph/0303020}

\combib{Dineen \& Coles }{2003}{}
\aut{Dineen}{P} \laut{Coles}{P}
\unrefs{}
{\mn}
{accepted}
{}
{2003}
{astro-ph/0306529}

\combib{Doroshkevich~\etal}{2003}{}
\aut{Doroshkevich}{A. G} \aut{Naselsky} {P. D} \aut{Verkhodanov} {O. V}
\aut {Novikov} {D. I} \aut{Turchaninov} {V. I} \aut{Novikov} {I. D}
\laut {Christensen} {P. R}
\unrefs{}
{\aaa}
{submitted}
{}
{2003}
{astro-ph/0305537}

\combib{Eriksen~\etal}{2003}{}
\aut{Eriksen}{H. K} \aut{Hansen} {F. K} \aut{Banday} {A. J}
\aut {G\'orski} {K. M} \laut{Lilje} {P. B}
\unrefs{}
{\apjl}
{submitted}
{}
{2003}
{astro-ph/0307507}

\combib{Fisher}{1993}{}
\laut{Fisher}{N. I}
\refs{}
{Statistical analysis of Circular Data}
{Cambridge University Press}
{Cambridge}
{1993}
{}

\combib{Gaztanaga~\etal}{2003}{}
\aut{Gaztanaga}{E} \aut{Wagg} {J} \aut{Multamaki} {T}
\aut {Montana} {A} \laut{Hughes} {D. H}
\unrefs{}
{\mn}
{accepted}
{}
{2003}
{astro-ph/0304178}

\combib{G\'{o}rski, Hivon \& Wandelt}{1999}{healpix}
\aut{G\'{o}rski} {K. M} \aut{Hivon} {E} \laut{Wandelt} {B. D}
\refs{}
{Proceedings
of the MPA/ESO Cosmology Conference ``Evolution of Large-Scale
Structure'', eds. A. J. Banday, R. S. Sheth and L. Da Costa}
{Print Partners Ipskamp}
{NL}
{1999}
{}


\combib{Hinshaw~\etal}{2003a}{}
\autetal{Hinshaw}{G}
% Komatsu, E.; Kogut, A.; Nolta, M. R.; Bennett, C. L.; Halpern, M.;
% Hinshaw, G.; Jarosik, N.; Limon, M.; Meyer, S. S.; Page, L.;
% Spergel, D. N.; Tucker, G. S.; Verde, L.; Wollack, E.; Wright, E. L.
\refs{}
{\apjs}
{148}
{63}
{2003}
{}

\combib{Hinshaw~\etal}{2003b}{}
\autetal{Hinshaw}{G}
%  Hinshaw, G.; Spergel, D. N.; Verde, L.; Hill, R. S.; Meyer, S. S.;
%  Barnes, C.; Bennett, C. L.; Halpern, M.; Jarosik, N.; Kogut, A.;
%  Komatsu, E.; Limon, M.; Page, L.; Tucker, G. S.; Weiland, J. L.;
%  Wollack, E.; Wright, E. L.
\refs{}
{\apjs}
{148}
{135}
{2003}
{}

\combib{Komatsu~\etal}{2003}{wmapng}
\autetal{Komatsu}{E}
% Komatsu, E.; Kogut, A.; Nolta, M. R.; Bennett, C. L.; Halpern, M.;
% Hinshaw, G.; Jarosik, N.; Limon, M.; Meyer, S. S.; Page, L.;
% Spergel, D. N.; Tucker, G. S.; Verde, L.; Wollack, E.; Wright, E. L.
\refs{}
{\apjs}
{148}
{119}
{2003}
{astro-ph/0203223}

\combib{Matsubara}{2003}{}
\laut{Matsubara}{T}
\refs{}
{\apj}
{591}
{L79}
{2003}
{astro-ph/0303278}

\combib{Naselsky\etal}{2003}{}
\aut{Naselsky}{P. D} \aut{Verkhodanov}{O. V} \aut{Chiang}{L.-Y}
\laut{Novikov}{I. D}
\unrefs{}
{\apj}
{submitted}
{}
{2003}
{astro-ph/0310235}

\combib{Naselsky\etal}{2002}{}
\aut{Naselsky}{P. D} \aut{Novikov}{D. I} \laut{Silk}{J}
\refs{}
{\mn}
{335}
{550}
{2002}
{}

\combib{Park}{2003}{park}
\aut{Park}{C.-G}
% Non-Gaussian Signatures in the Temperature Fluctuation Observed by the WMAP
\unrefs{}
{\mn}
{submitted}
{}
{2003}
{astro-ph/0307469}
{}


\combib{Seljak \& Zaldarriaga}{1996}{}
\aut{Seljak}{U} \laut{Zaldarriaga}{M}
\refs{}
{\apj}
{469}
{437}
{1996}
{astro-ph/9603033}

\combib{Tegmark, de Oliveira-Costa \& Hamilton}{2003}{tegmark} 
\aut{Tegmark}{M} \aut{de Oliveira-Costa}{A} \laut{Hamilton}{A} 
\unrefs{}
{\prd}
{submitted}
{}
{2003}
{astro-ph/03022496}

\combib{Vielva\etal}{2003}{}
\aut{Vielva}{P} \aut{Martinez-Gonzalez}{E} \aut{Barreiro}{R. B}
\aut{Sanz}{J. L} \laut{Cayon}{L}
\unrefs{}
{}
{submitted}
{}
{2003}
{astro-ph/0310273}


\end{thebibliography}
\end{document}